\documentclass[journal]{IEEEtran}

\usepackage{amsfonts}
\usepackage{amsmath}
\usepackage{amssymb}
\usepackage{mathabx}
\usepackage{mathrsfs}
\usepackage{tensor}
\usepackage{esint}
\usepackage{tikz}
\usepackage{multirow}
\usepackage{makecell}
\usepackage{textcomp}

\newcolumntype{C}[1]{>{\centering\arraybackslash}m{#1}}

\DeclareMathAlphabet{\mathpzc}{OT1}{pzc}{m}{it}
\newcommand{\Px}{\mbox{\textbf{P}}}
\newcommand{\Py}{\mbox{$\textbf{P}_{\star}$}}
\newcommand{\x}{\mbox{\textbf{x}}}
\newcommand{\y}{\mbox{\textbf{y}}}
\newcommand{\OS}{\mbox{\textbf{O}}}

\begin{document}
\title{Metal Artifact Reduction with Intra-Oral Scan Data for 3D Low Dose Maxillofacial CBCT Modeling}
	
\author{\IEEEauthorblockN{Chang Min Hyun\IEEEauthorrefmark{1}, Taigyntuya Bayaraa\IEEEauthorrefmark{1}, Hye Sun Yun\IEEEauthorrefmark{1}, Tae-Jun Jang\IEEEauthorrefmark{1}, Hyoung Suk Park\IEEEauthorrefmark{2}, and Jin Keun Seo\IEEEauthorrefmark{1}}
	\\ \IEEEauthorblockA{\IEEEauthorrefmark{1}School of Mathematics and Computing (Computational Science and Engineering), Yonsei University, Seoul, 03722, South Korea} \\ \IEEEauthorblockA{\IEEEauthorrefmark{2}National Institute for Mathematical Sciences, Daejeon, 34047, South Korea}
	\thanks{Manuscript received XXX; revised  XXX. Corresponding author: Hyoung Suk Park (hspark@nims.re.kr).}}
	
% The paper headers
\markboth{}%
	{ \MakeLowercase{\textit{C. M. Hyun et al.}}: }
	
\IEEEtitleabstractindextext{%
		
\begin{abstract}
Low-dose dental cone beam computed tomography (CBCT) has been increasingly used for maxillofacial modeling. However, the presence of metallic inserts, such as implants, crowns, and dental filling, causes severe streaking and shading artifacts in a CBCT image and loss of the morphological structures of the teeth, which consequently prevents accurate segmentation of bones. A two-stage metal artifact reduction method is proposed for accurate 3D low-dose maxillofacial CBCT modeling, where a key idea is to utilize explicit tooth shape prior information from intra-oral scan data whose acquisition does not require any extra radiation exposure. In the first stage, an image-to-image deep learning network is employed to mitigate metal-related artifacts. To improve the learning ability, the proposed network is designed to take advantage of the intra-oral scan data as side-inputs and perform multi-task learning of auxiliary tooth segmentation. In the second stage, a 3D maxillofacial model is constructed by segmenting the bones from the dental CBCT image corrected in the first stage. For accurate bone segmentation, weighted thresholding is applied, wherein the weighting region is determined depending on the geometry of the intra-oral scan data. Because acquiring a paired training dataset of metal-artifact-free and metal artifact-affected dental CBCT images is challenging in clinical practice, an automatic method of generating a realistic dataset according to the CBCT physics model is introduced. Numerical simulations and clinical experiments show the feasibility of the proposed method, which takes advantage of tooth surface information from intra-oral scan data in 3D low dose maxillofacial CBCT modeling.
\end{abstract}

}
\maketitle
\IEEEdisplaynontitleabstractindextext
\IEEEpeerreviewmaketitle

\section{Introduction}
\begin{figure*}[t]
	\centering
	\includegraphics[width=0.95\textwidth]{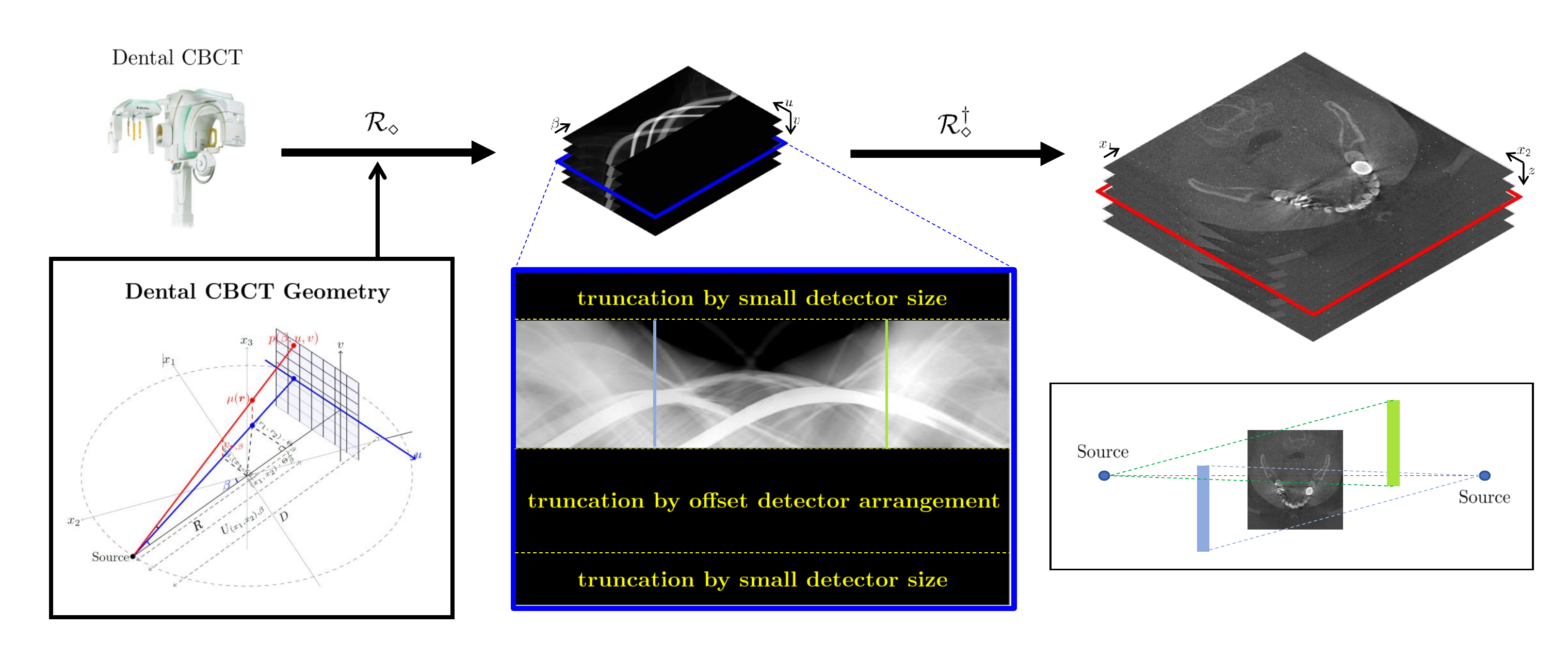}
	\caption{Low-dose dental CBCT uses a small detector with offset array. The small detector size leads to a small area of the scanner FOV, which causes the patient's head to be cut off the sinogram data in the transversal direction. This incomplete sinogram data can be combined with beam hardening of the teeth, creating streaked artifacts. Photon starvation is very common in dental low-dose X-ray CBCT, especially when the patient has many implants.}
	\label{intro}
\end{figure*}

In the field of dentistry, dental cone beam computed tomography (CBCT)-based maxillofacial modeling has been widely utilized to understand the complicated anatomical structures of the mandible, maxilla, or skeleton for various clinical tasks \cite{gupta2013,marchetti2007,miracle2009,sukovic2003,swennen2009,Swennen2009_2,scarfe2017,weiss2019}. Dental CBCT provides a three-dimensional (3D) maxillofacial model with a reasonably high resolution at a low radiation dose and cost. However, in the presence of metallic inserts, such as implants, crowns, and dental filling, metal artifacts cause the reconstructed CBCT image to deteriorate, making it difficult to segment tooth structures accurately \cite{gateno2007,nardi2015,santler1998,schulze2011}. The metal artifacts are caused by several physical factors, such as beam hardening, scattering, noise, nonlinear partial volume, and photon starvation. Reducing metal-related artifacts in low-dose dental CBCT has drawn increased attention in digital dentistry workflows because the number of people with oral metallic appliances is rapidly and steadily growing \cite{draenert2007,esmaeili2012,pauwels2013,razavi2010,sanders2007,schulze2010,sancho2015}.

There have been numerous studies on metal artifact reduction (MAR) in computed tomography (CT) imaging, which include sinogram inpainting-based correction \cite{Abdoli2010,Bazalova2007,Kalender1987,Meyer2010,Park2013}, statistical iterative correction \cite{DeMan2001,Elbakri2002,menvielle2006,OSullivan2007,Williamson2002}, and dual-energy reconstruction approaches \cite{Alvarez1976,Lehmann1981,Yu2012}. However, existing MAR methods are not fully satisfactory for clinical use. The inpainting-based correction approach can generate secondary artifacts owing to inaccurate interpolation along the metal trace in the sinogram. The statistical iterative correction and dual-energy approaches require a large computational cost and an additional radiation dose, respectively.

Recent advances in deep learning technology are progressing in MAR. \cite{zhang2018} used deep learning to generate an artifact-reduced prior image, then used the projection of the prior image to replace the metal-affected projection, and then performs the final MAR CT reconstruction. \cite{gjesteby2019} performed a residual learning to correct metal streaking artifacts after the first pass by normalized metal artifact reduction (NMAR). Metal artifacts are non-local and highly associated with various factors, including the geometry of metallic inserts and the energy spectrum of the incident X-ray beam, making it difficult to learn their complicated structures in the image domain.

\cite{park2018} proposed a deep learning-based sinogram correction method to reduce the primary metal-induced beam-hardening factors along the metal trace in the sinogram. This method was applied in the restricted situation of a patient-implant-specific model in which two simple metallic objects were placed in the hip joint. \cite{lin2019} proposed a dual-domain network (DuDoNet) to restore sinogram consistency and enhance CT images simultaneously. DuDoNet pursues MAR enhancement by leveraging dual-domain learning networks: sinogram and image enhancement networks. Here, the dual networks operate separately on the sinogram and image domains, but are trained in an end-to-end fashion. The sinogram enhancement network learns how to correct a metal-affected sinogram by using an inpainting loss on the metal trace and sinogram consistency loss.  Since the final image is the output of the image enhancement network, the data fidelity may be compromised, resulting in anatomical structure changes. \cite{yu2020} proposed a dual-domain joint learning network that first generates a good prior image with fewer metal artifacts; then, the forward projection of the prior image is utilized for sinogram enhancement. The final output is the reconstructed image from the sinogram modified only in the metal trace area.

Although the above sinogram-domain learning methods, including dual-domain learning, have shown potential to improve the overall image quality, it is difficult to apply such approaches to a practical dental CBCT environment in terms of offset detection, field of view (FOV) truncation, low radiation dose, and 3D characteristics in image reconstruction.
First, dental CBCT sinogram data are severely contaminated under the influence of complex factors related to complex geometries of teeth and bones, FOV truncation, offset detection, scattering, etc. See Fig. \ref{intro}. Hence, sinogram inpainting using deep learning is much more difficult than that in a standard CT environment. Sinogram inpainting is very different from image inpainting, because sinogram inpainting requires sophisticated utilization of global information outside the inpainting area, as well as surrounding local information, while maintaining sinogram consistency. Second, in 3D CT, sinogram learning or even simultaneous learning of image and sinogram domain networks is challenging because of several factors, such as high dimensionality, and huge computation and memory burdens. The aforementioned MAR methods, including sinogram-domain learning, have focused on a 2D fan-beam CT environment. Most importantly, dental CBCT scans are obtained with a significantly lower radiation dose. Therefore, when multiple and strong metallic inserts occupy a significant area, the corresponding sinogram is frequently missing along the metal trace, resulting in the loss of tooth structures around the metallic inserts in the reconstructed image. Unfortunately, it is still arduous to restore the missing morphological structures effectively, regardless of the remarkable capability deep learning has shown for estimating expected values by exploring prior information from the training dataset \cite{hyun2021,seo2019}.

\begin{figure}[t]
	\centering
	\includegraphics[width=0.475\textwidth]{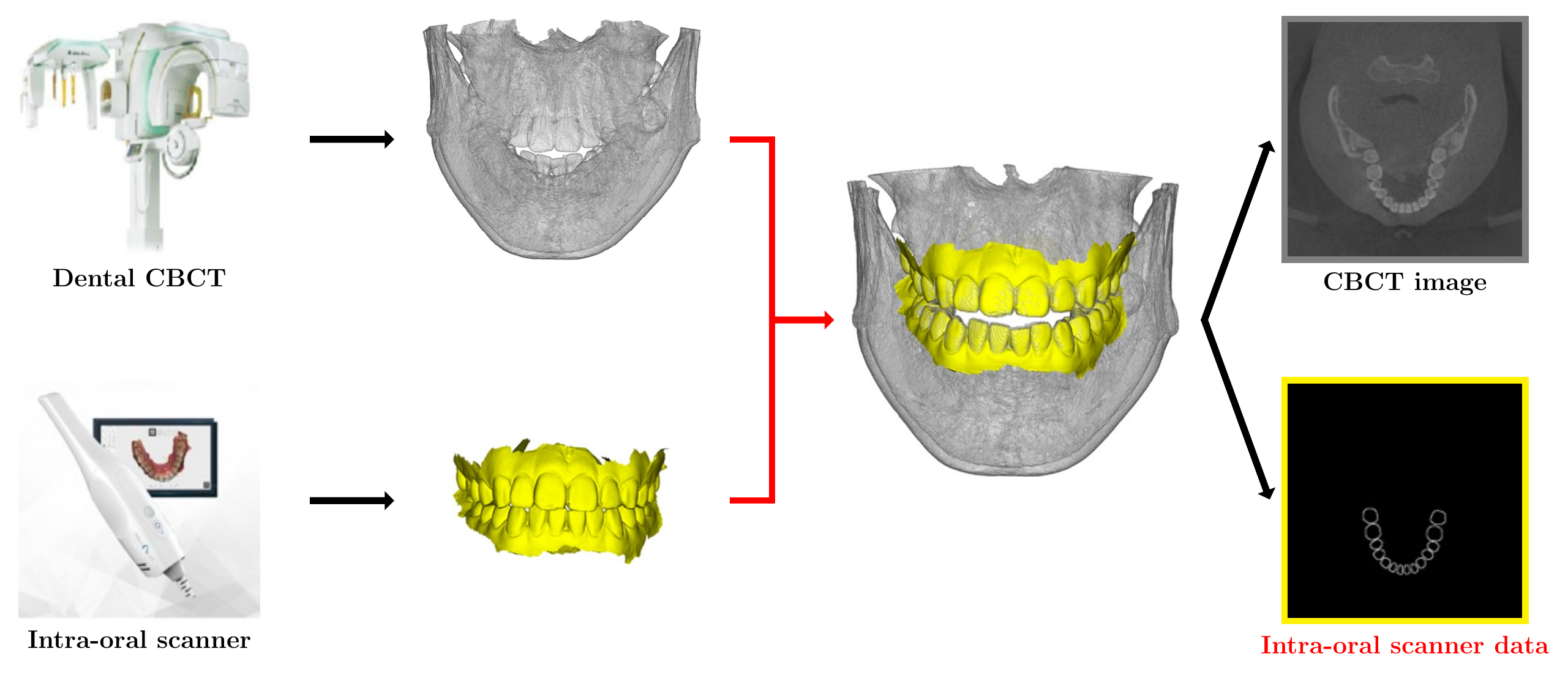}
	\caption{3D dental CBCT and intra-oral scan data. The intra-oral scan data can provide 3D surface information of teeth. We assume that intra-oral scanning provides exact tooth boundary information.}
	\label{oralscandata}
\end{figure}

With the development of intra-oral scanning technology, intra-oral scanners are rapidly being adopted in workflows of digital dentistry \cite{park2015}. As shown in Fig. \ref{oralscandata}, the intra-oral scanner makes it possible to acquire 3D surface information of teeth directly from the oral cavity, and a joint utilization of intra-oral scan with CBCT has been being considered for several clinical applications (e.g., orthognathic surgery planning, implant guiding, and occlusion analysis) that require sophisticated understanding of both maxillofacial structure and dentition \cite{albayrak2021,flugge2017,gateno2003,lee2021,plooij2011,rangel2018,roig2020,shujaat2021}.

\begin{figure*}[t]
	\centering
	\includegraphics[width=0.975\textwidth]{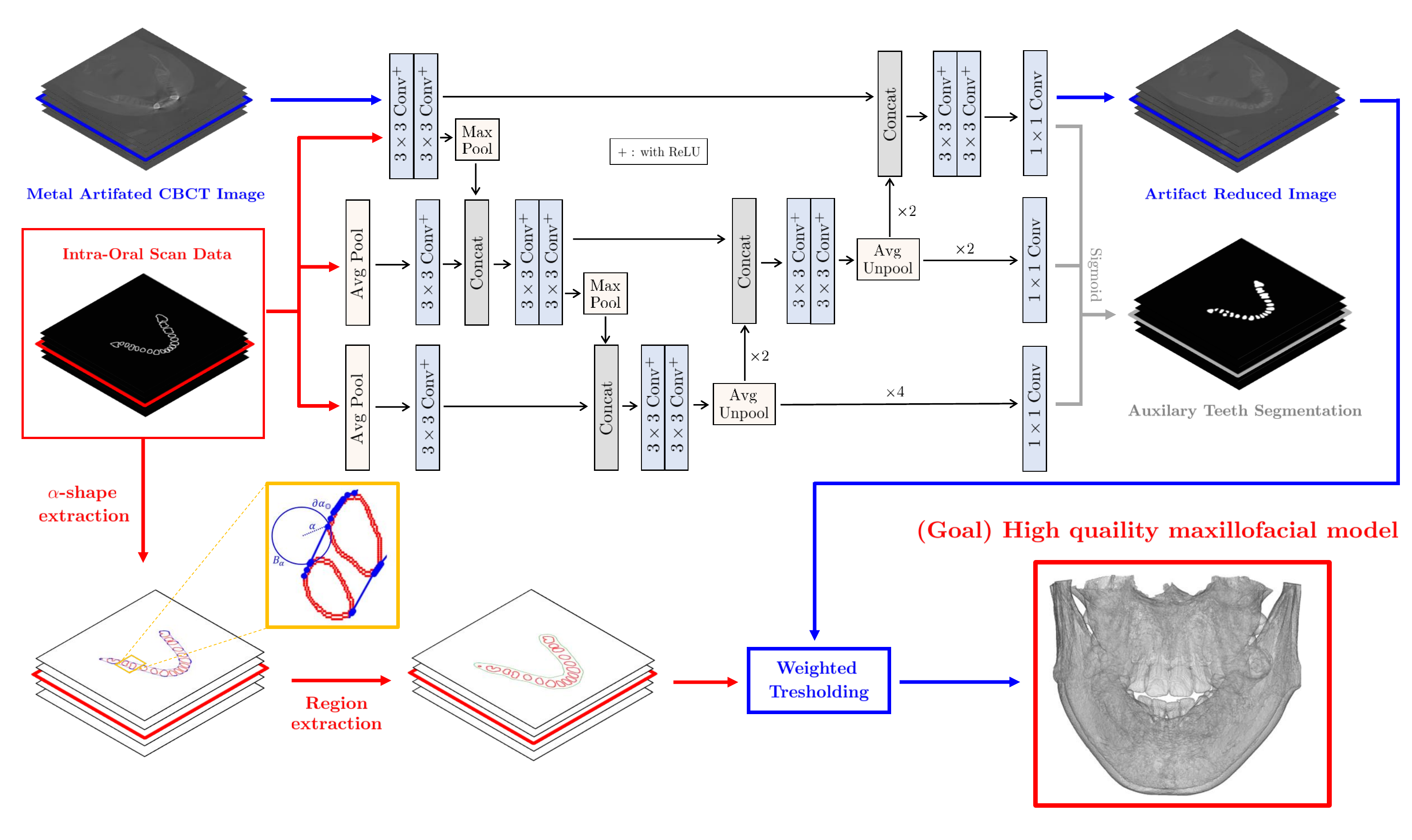}
	\caption{Overall process of the proposed metal artifact reduction method with explicit shape-prior of intra-oral scan data for 3D low dose maxillofacial CBCT imaging.}
	\label{proposedmethod}
\end{figure*}

In this study, a novel MAR method is developed to construct a 3D maxillofacial model from low-dose dental CBCT, where a key idea is to utilize explicit tooth shape prior information from intra-oral scan data whose acquisition does not require any extra radiation exposure. The proposed method comprises two core stages; (i) deep learning-based MAR and (ii) maxillofacial modeling. In the first stage, image-to-image deep learning is employed to mitigate metal-related artifacts from CBCT images, where metal-artifacted and  images are used as input and ground-truth, respectively. To improve the learning ability, the proposed network is designed to take advantage of the intra-oral scan data as side-inputs and perform the multi-task learning of auxiliary tooth segmentation \cite{caruana1997,sun2019}. The suitable incorporation with explicit shape-prior of intra-oral scan data can bring significant positive effect in terms of learnability and feature extraction \cite{liu2021}. In the second stage, a 3D maxillofacial model is constructed by segmenting the bones from the CBCT images corrected in the previous stage. For more accurate modeling, the weighted thresholding is applied, where the weight region is determined depending on the geometry of the intra-oral scan data.

In clinical practice, it is challenging to obtain a paired training dataset of  and metal-artifacted CBCT images and corresponding intra-oral scan data, simultaneously. To circumvent this difficulty, a realistic paired training dataset for MAR is generated through the following procedures, which do not involve any time-consuming and labor-intensive manual process. Metallic inserts, such as dental crowns and tooth implants, are automatically generated in basis of individual tooth segmentation from a metal-artifact-free CBCT image. Then, according to the CBCT physics model, realistic metal-affected sinogram data are generated. For intra-oral scan data generation, it is assumed that intra-oral scanning provides exact tooth boundary information in CT images.

This study, for the first time, investigates the potential impact of using intra-oral scan data in MAR and maxillofacial modeling. Numerical simulations and clinical experiments demonstrated the feasibility of the proposed method and the benefit of using intra-oral scan data in 3D low dose maxillofacial CBCT imaging.

\section{Method} \label{Method}
The proposed method is developed for 3D low-dose maxillofacial CBCT imaging, where the sinogram data $\Px$ can be expressed as
\begin{equation}\label{Beerslaw}
	\Px = \mathcal S_{\mbox{\footnotesize ub}}(-\mbox{ln} \int_{E} \eta(E)\exp(-\mathcal R_{\diamond} \mu_{E}) dE + \textbf{n})
\end{equation}
Here, $\mu_{E}$ is the attenuation coefficient distribution of a 3D human body to be scanned at an energy $E$, $\eta$ is the normalized energy distribution of the X-ray source, $\textbf{n}$ is the CT noise, and  $\mathcal S_{\mbox{\footnotesize ub}}$ is a subsampling operator determined by the size and arrangement of the detector (typically, small and offset). In the presence of metallic objects inside the FOV, the standard FDK algorithm \cite{feldkamp1984}, denoted by $\mathcal R^{\dagger}_{\diamond}$, produces severe streaking and shadowing artifacts that cause the image quality of maxillofacial structures to deteriorate. Hence, high-quality 3D maxillofacial imaging is arduous only with the image $\mathcal R^{\dagger}_{\diamond}\Px$.

The goal of the proposed method is to provide a high-quality 3D maxillofacial image from metal-affected sinogram data $\Px$ by leveraging intra-oral scan data $\OS$. The output should be competitive with a ``gold-standard" maxillofacial image $\y_{\mbox{\footnotesize mf}}$ acquired from an artifact-free CT image $\y = \mathcal R^{\dagger}_{\diamond} \Py$, where $\Py$ represents the artifact-free sinogram data corresponding to $\Px$. 

The intra-oral scan data $\OS$ provide a 3D tooth surfaces that can be useful as prior information about tooth geometry. It is assumed that intra-oral scan data $\OS$ provides exact tooth boundary information.

The proposed method is based on the image-to-image learning approach and weighted thresholding that leverages intra-oral scan data as explicit shape prior information of tooth geometry for MAR. As illustrated in Fig. \ref{proposedmethod}, the reconstruction map $f$ can be expressed as
\begin{equation} \label{entire_map}
	f = f_{\mbox{\footnotesize{$\alpha$-WT}}} \circ f_{\mbox{\footnotesize{IE}}} \circ \mathcal R^{\dagger}_{\diamond}
\end{equation}
where
\begin{itemize}
	\item $\mathcal R^{\dagger}_{\diamond}$ is the weighted FDK algorithm involving the sinogram extrapolation method for addressing offset detector arrangement and FOV truncation.
	\item $f_{\mbox{\footnotesize{IE}}}$ is the tooth geometry prior information-based-image-enhancing network $f_{\mbox{\footnotesize{IE}}}$, which mitigates metal-related artifacts.
	\item $f_{\mbox{\footnotesize{$\alpha$-WT}}}$ is a weighted thresholding, wherein the weighting region is determined in basis of the $\alpha$-shape from intra-oral scan data. This procedure is used for further removing the remaining streaking artifacts around the teeth in constructing a maxillofacial model.
\end{itemize}
Here, the input of $f$ is a pair of metal-affected data $\Px$ and intra-oral scan data $\OS$ (i.e., $f : (\Px,\OS) \mapsto f(\Px,\OS)\approx \textbf y_{\mbox{\scriptsize mf}}$).

The FDK reconstruction $\mathcal R_{\diamond}^{\dagger}$ is as follows. For given $\mbox{\textbf{P}}$, an image $\mathcal R_{\diamond}^{\dagger} \mbox{\textbf{P}}(\boldsymbol{x},z)$ is defined by
\begin{equation}
	\int_{0}^{2\pi} \omega(u_{\beta,\textbf{x}}) \int_{\mathbb{R}} \frac{\mathcal P (\mbox{\textbf{P})}(\beta,u,v_{\beta,\boldsymbol{x},z})R^3\hbar(u_{\beta,\boldsymbol{x}}-u)}{4\pi U_{\beta,\boldsymbol{x}}^2 \sqrt{R^2+u^2+v_{\beta,\boldsymbol{x},z}^2}}dud\beta
\end{equation}
where $\omega$ is a weighting function satisfying $\omega(u)+\omega(-u)=1$ and $\mathcal P$ is a constant padding operator that fills the truncated regions by boundary values \cite{bayaraa2020,sharma2014,tisson2006}. Here, $(\boldsymbol{x}, z)$ with $\boldsymbol{x} = (x_1,x_2)$ is the 3D Cartesian coordinate system, $\beta$ is the projection angle of the X-ray source rotated along the circular trajectory, $(u,v)$ is the coordinate system of the 2D flat-panel detector, $R$ is the distance from the X-ray source to the isocenter, $\hbar$ is the 1D ramp filter, $U_{\beta,\boldsymbol{x}}=R+\boldsymbol{x}\cdot\theta_{\beta}^\perp$, $v_{\beta,\boldsymbol{x},z}=zR/U_{\beta,\boldsymbol{x}}$, $u_{\beta,\boldsymbol{x}}=R(\boldsymbol{x}\cdot\theta_{\beta})/U_{\beta,\boldsymbol{x}}$, $\theta_{\beta}=(\cos \beta, \sin \beta)$, and $\theta_{\beta}^\perp=(-\sin \beta, \cos \beta)$. See the blue box in Fig. \ref{intro}.

\subsection*{Stage 1. Image-enhancing network $f_{\mbox{\footnotesize{IE}}}$}
The image-enhancing network $f_{\mbox{\footnotesize{IE}}}$ is designed to utilize intra-oral scan data as the tooth shape prior while mitigating metal-related artifacts. In our experience, an image domain-learning-based approach can mitigate metal-related artifacts effectively, whereas it tends to have weakness in recovering tooth shape. especially when being destroyed by severe artifacts or when being missed. To compensate for this weakness, we attempt to take advantage of supplemental shape information from intra-oral scan data. We emphasize that data acquisition by the intra-oral scanner does not increase the total amount of radiation exposure to a patient.

Let $\x$ be a 3D CBCT image reconstructed using the FDK algorithm (i.e., $\x=\mathcal R^{-1}_{\diamond}(\Px)$). The image-enhancing network $f_{\mbox{\footnotesize{IE}}}$ aims to provide
\begin{equation} \label{objec1}
	f_{\mbox{\footnotesize{IE}}}(\x_j,\OS_j) \approx \y_j
\end{equation}
where $\y_j$ is the $j$-th slice of a metal=artifact-free image (i.e., $\y$ = $ \mathcal R^{-1}_{\diamond} \textbf P_{\star}$). It is also desirable that the output satisfies
\begin{equation} \label{objec2}
	\partial f_{\mbox{\footnotesize{IE}}}(\x_j,\OS_j) |_{\mbox{\scriptsize teeth}} \approx \OS_j
\end{equation}
where $\partial f_{\mbox{\footnotesize{IE}}}(\x_j,\OS_j;\textbf{w}_2) |_{\mbox{\scriptsize teeth}}$ is a binary mask of the tooth surface region on the output image $f_{\mbox{\footnotesize{IE}}}(\x_j,\OS_j;\textbf{w}_2)$.

To accomplish these goals, two strategies are adopted; side-input layer and multi-task learning. First, additional information of intra-oral scan data is repeatedly enriched during feature extraction in an encoding path. These side inputs can help the network extract tooth shape while compensating for missing or severely distrusted structures through high quality shape information provided by intra-oral scan data. Second, multi-task learning is applied, which learns image reconstruction and auxiliary tooth segmentation in a parallel fashion. In the medical imaging field, it has been reported that deep learning-based image reconstruction ability can be boosted by learning other image-related tasks, such as segmentation and registration \cite{liu2021,sun2019}. In terms of image recovery, the auxiliary tooth segmentation is expected to reveal the shapes of the teeth in the decoding path and the interference of tooth features, which are joint domain information of the interrelated tasks, through the shared parameters.

Fig. \ref{proposedmethod} shows the overall procedure of the proposed image-enhancing network $f_{\mbox{\footnotesize{IE}}}$. Inspired by M-net \cite{mehta2017}, the proposed network has side-input and side-output layers. In the side-input layers, intra-oral scan data $\OS$ with suitable resizing is repeatedly added to the encoding path after $3\times 3$ convolution. In the side-output layers, tooth segmentation masks are obtained during the decoding path. The detailed backbone structure can be found in \cite{mehta2017}.

When $(\textbf{s}_0)^{(i)}_{j}$ is the final network output of $i$-th training data and $j$-th slice (i.e., $(\textbf{s}_0)^{(i)}_{j}=f_{\mbox{\footnotesize{IE}}}(\x_j^{(i)}, \OS_j^{(i)})$), the network $f_{\mbox{\footnotesize{IE}}}$ is trained as follows.
\begin{align}
	\underset{f}{\mbox{argmin}} & \sum_{i} ~ \mathcal L_{\ell_2}( ((\textbf{s}_0)^{(i)}_{j})^{(1)},\textbf \y_j^{(i)}) \nonumber \\ & +  \mathcal L_{ce}( ( (\textbf s_0)^{(i)}_j)^{(2)} ,\textbf{S}_j^{(i)}) + \sum_{k=1}^{2} \mathcal L_{ce}( ( \textbf s_k)^{(i)}_j ,\textbf{S}_j^{(i)}) 
\end{align}
where $((\textbf{s}_0)^{(i)}_{j})^{(1)}$ denotes the first channel output of $(\textbf{s}_0)^{(i)}_{j}$, $\textbf{S}^{(i)}_j$ denotes a binary segmentation mask of the reference tooth region in the metal-artifact-free image $\y^{(i)}_j$, $\{(\textbf{s}_k)^{(i)}_{j}\}_{k=1}^{2}$ is a set of side outputs in the decoding path, $\mathcal L_{\ell_2}$ is the standard $\ell_2$ loss, and $\mathcal L_{ce}$ is the cross-entropy loss. For convenience, the notation $f_{\mbox{\footnotesize{IE}}}(\x,\OS)$ is used to represent the output image (i.e., the first channel output).

\begin{figure}[h!]
	\centering
	\includegraphics[width=0.475\textwidth]{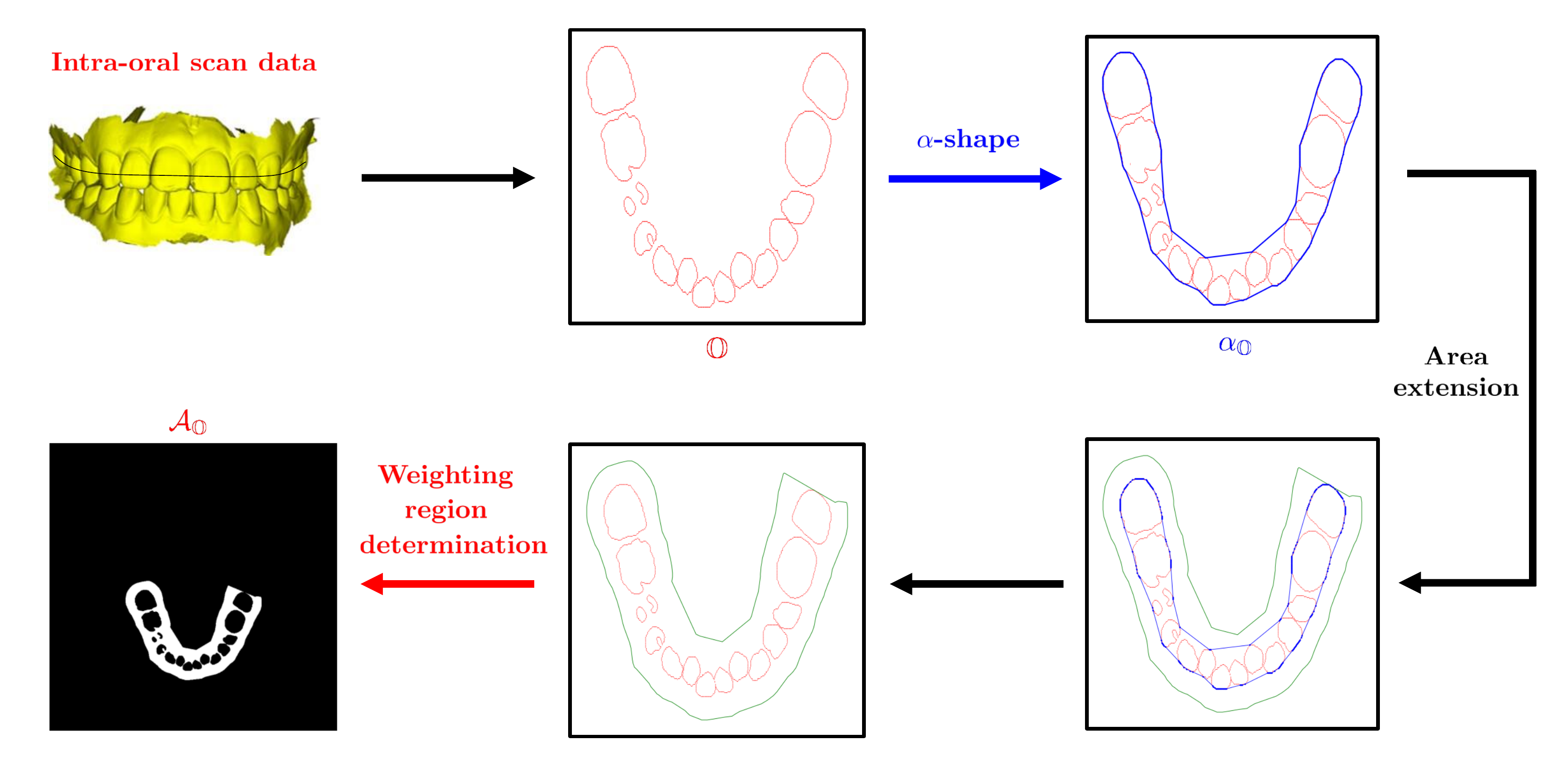}
	\caption{$\alpha$-shape-based region determination for weighted thresholding $f_{\mbox{\footnotesize{$\alpha$-WT}}}$}
	\label{alphashape}
\end{figure}
\subsection*{Stage 2. $\alpha$-shape-based weighted thresholding $f_{\mbox{\footnotesize{$\alpha$-WT}}}$}	\label{part3method}
The next step is maxillofacial imaging from the metal-artifact-reduced CBCT image obtained in the previous stage. A final 3D maxillofacial image is obtained by weighted thresholding, which can further reduce the remaining streaking artifacts around teeth. The weighting region is determined depending on the geometry of the intra-oral scan data $\OS$. To extract the geometry, the $\alpha$-shape technique \cite{edelsbrunner1994} is used. It provides a family of piece-wise linear lines associated with the shape of the teeth. Fig. \ref{alphashape} shows the overall process.

When $\y_{\mbox{\footnotesize dl}}$ is $\y_{\mbox{\footnotesize dl}} = f_{\mbox{\footnotesize IE}}(\x,\OS)$, the weight thresholding $f_{\mbox{\footnotesize{$\alpha$-WT}}}$ can be expressed as
\begin{equation} \label{wt_maxillo}
	f_{\mbox{\footnotesize{$\alpha$-WT}}}^\tau(\y_{\mbox{\footnotesize dl}},\OS) = \widehat{\y}_{\mbox{\footnotesize mf}}
\end{equation}
where
\begin{equation}
	\widehat{\y}_{\mbox{\footnotesize mf}} =
	\left\{\begin{array}{cl}
		\widehat{\y}_{\mbox{\footnotesize mf}}(p) = 0 & \mbox{if }p \in \mathcal A_{\mbox{\footnotesize \OS}} \mbox{ or } \y_{\mbox{\footnotesize mf}}(p) < \tau \\
		\widehat{\y}_{\mbox{\footnotesize mf}}(p) = 1 & \mbox{otherwise } \\
	\end{array}\right.
\end{equation}
Here, $p$ is a point in a grid of $\y_{\mbox{\footnotesize mf}}$, $\tau$ is a thresholding constant, and $\mathcal A_{\mbox{\footnotesize \OS}}$ is a thresholding region obtained using the $\alpha$-shape from $\OS$.

The region  $\mathcal A_{\mbox{\footnotesize \OS}}$ is obtained as follows. For given intra-oral scan data $\OS$, $\mathbb{O}$ is a point cloud corresponding to $\OS$. Denoted by $\alpha_{\footnotesize \mathbb{O}}$, an $\alpha$-shape of $\mathbb{O}$ is given by a polytope with a boundary $\partial \alpha_{\footnotesize \mathbb{O}}$, which is defined by
\begin{equation}
	\partial \alpha_{\footnotesize \mathbb{O}} =  \{ ~  \Delta_\mathbb{T} ~ | ~ \mathbb{T} \subset \mathbb{O}, |\mathbb{T}|\leq 3,\Delta_\mathbb{T} \mbox{ is } \alpha\mbox{-exposed} ~ \}
\end{equation}
where $\Delta_\mathbb{T}$ denotes a simplex for $\mathbb{T}$, and $\Delta_\mathbb{T}$ is $\alpha$-exposed if and only if there exists an open ball $B_{\alpha}$ with radius $\alpha$ such that $B_{\alpha} \cap \mathbb{O}=\emptyset$ and $\partial B_{\alpha} \cap \mathbb{O} = \mathbb{T}$. Here, $\partial B_{\alpha}$ is a boundary of $B_{\alpha}$. After the $\alpha$-shape is obtained, an extension direction on each vertex of $\alpha_{\footnotesize \mathbb{O}}$ is defined by taking the average of the normal vectors on the faces that contain the vertex. Along the direction, $\alpha_{\footnotesize \mathbb{O}}$ is extended while preserving its shape and converted into a binary mask $\alpha_{\footnotesize \textbf{O}}$, where the inner regions of the shape boundary are filled with one. Finally, the region $\mathcal A_{\mbox{\footnotesize \OS}}$ is determined by
\begin{equation}
	\mathcal A_{\mbox{\footnotesize \OS}} = \alpha_{\footnotesize \textbf{O}} - \overline{\OS}
\end{equation}
where $\overline{\OS}$ is the binary mask where the inner part of tooth surfaces in $\OS$ is filled with one.

\section{Experiment and Result} \label{Result}
\begin{figure*}[t!]
	\centering
	\includegraphics[width=0.975\textwidth]{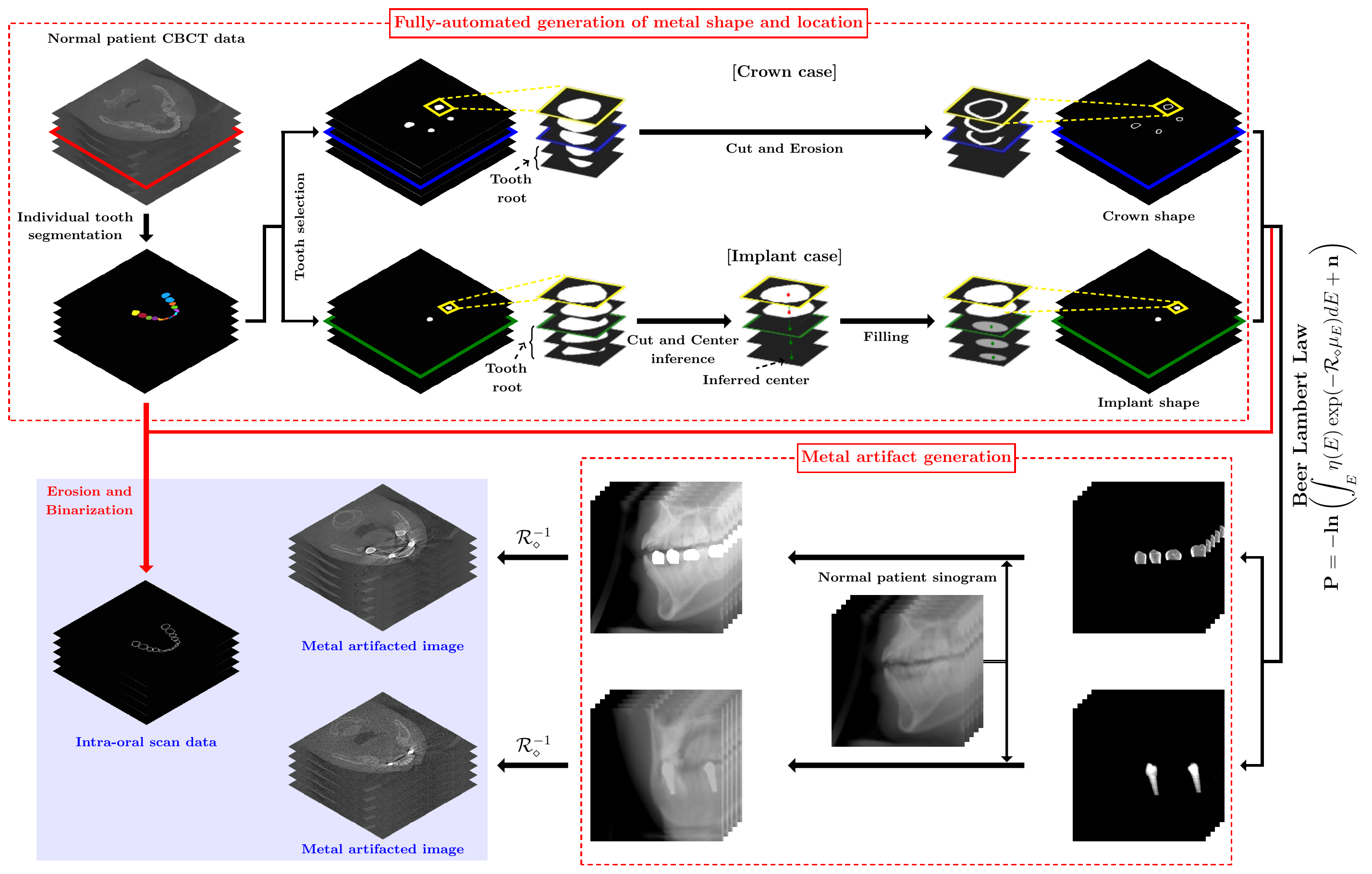}
	\caption{Overall process of fully-automated paired training data generation}
	\label{trainingdatageneration}
\end{figure*}
\subsection{Experiment Setting}
The sinogram data of a real patient were obtained from a commercial CBCT machine (Q-FACE, HDXWILL).
The voxel size was $1200\times654\times658$ with real scale of $0.2$ mm for each axis, where $1200$ is the number of uniformly sampled projection views in $[0,2\pi)$, and $654 \times 658$ is the number of samples measured by the 2D flat detector for each projection view. CBCT images were reconstructed in a voxel size of $800\times800\times400$ with a real scale of $0.2$mm. For cone beam projection and back-projection (FDK reconstruction), an open-source code, TIGRE \cite{biguri2016}, was used after modification to make it suitable for the experimental setting.

All simulated data were consistently generated to have same scale as the real data. A self-developed fully-automated paired data generation tool was used. The detailed process is described in Section \ref{datageneration}.

Metal-free CBCT sinogram data were collected from 20 normal patients. They were used for training data generation. Metal-affected CBCT data were collected from nine patients. They were used for test purposes. Among the metal-affected data, real intra-oral scan data for one patient was provided. The intra-oral scan data was acquired from a scanner (i500, MEDIT), where the file format was provided by the Standard Triangle Language (STL). A set of its vertices is a point cloud in millimeters, where the maxilla and mandible are represented by approximately 100,000 and 70,000 points, respectively. For registration into the dental CBCT system, the method described by \cite{jang2021_2} was applied. 

In PyTorch environment \cite{paszke2019}, all deep learning experiments were conducted with a computer system equipped with two Intel Xeon CPUs E5-2630 v4, 128GB DDR4 RAM, and four NVIDIA GeForce GTX 2080ti GPUs. The optimization was conducted using Adam optimizer \cite{kingma2014} and multi-GPUs. Batch normalization was applied to achieve fast convergence and minimization \cite{ioffe2015}. The network capacity (i.e., feature and network depths) was minimized as much as possible while maintaining the backbone structure because of the huge computational cost associated with the CBCT image size of $800\times800\times400$.

For $\alpha$-shape implementation, open source packages, Visualization ToolKit (VTK) and Alpha Shape Toolbox (AST), were used. The adaptive values $\alpha$ and $\tau$ were selected empirically.

\begin{figure*}[t!]
	\centering
	\includegraphics[width=0.975\textwidth]{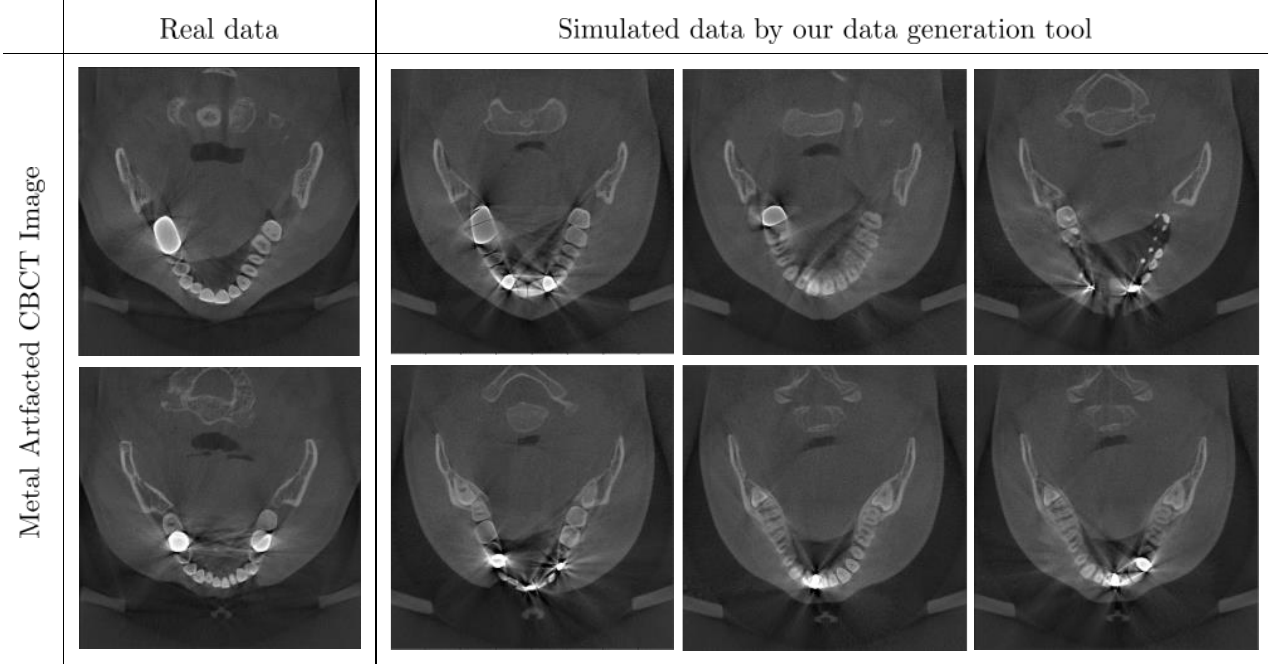}
	\caption{Real and simulated CBCT data. The simulated data is generated by the fully-automated data generation tool represented in Section \ref{datageneration}.}
	\label{trainingdatageneration2}
\end{figure*}
\subsection{Fully-Automated Paired Data Generation} \label{datageneration}
\begin{figure*}[t!]
	\centering
	\includegraphics[width=0.975\textwidth]{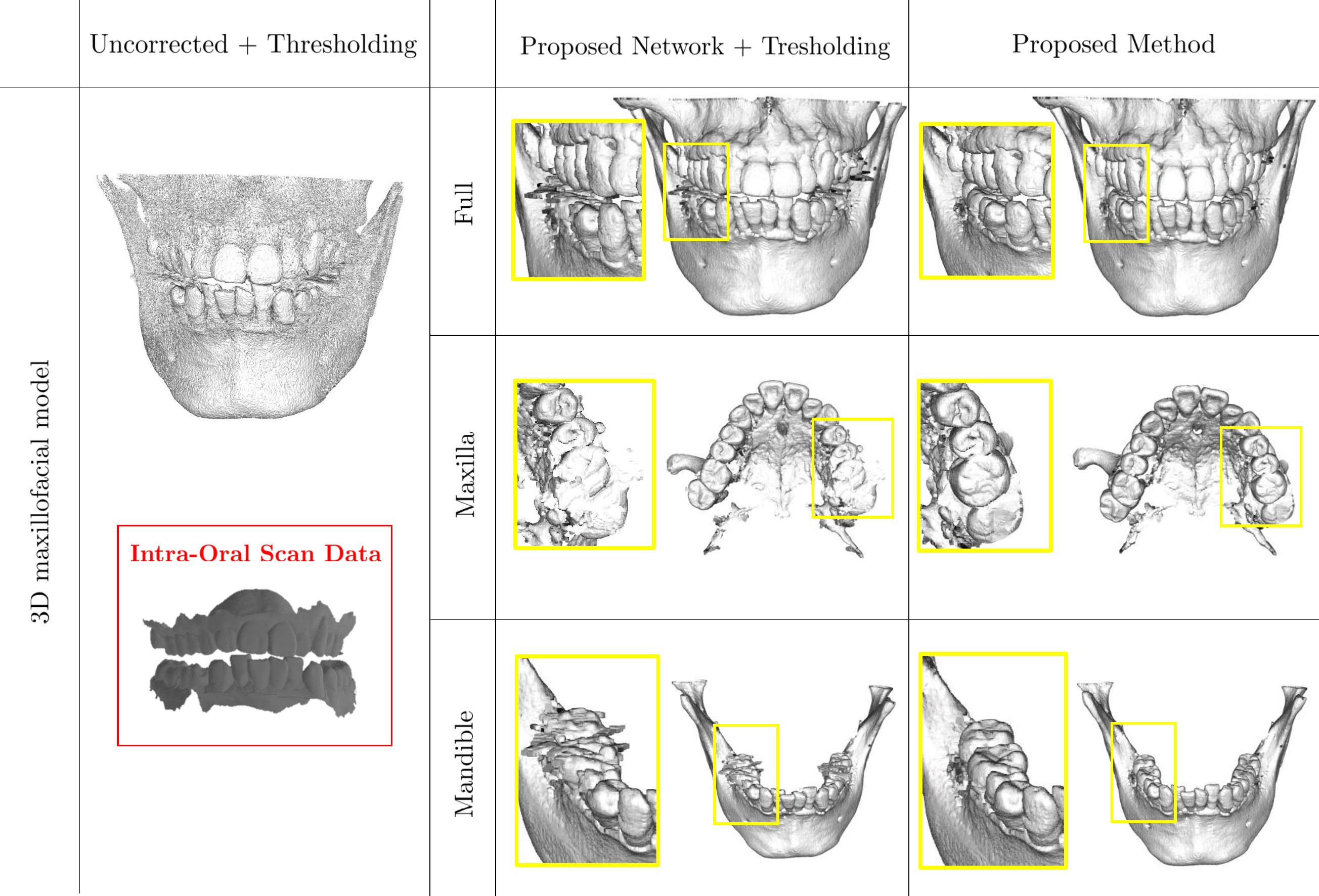}
	\caption{CBCT-based 3D maxillofacial modelling via the proposed method with clinical CBCT and real intra-oral scan data.}
	\label{exp_result_real_maxillofacial}
\end{figure*}
\begin{figure*}[t!]
	\centering
	\includegraphics[width=0.975\textwidth]{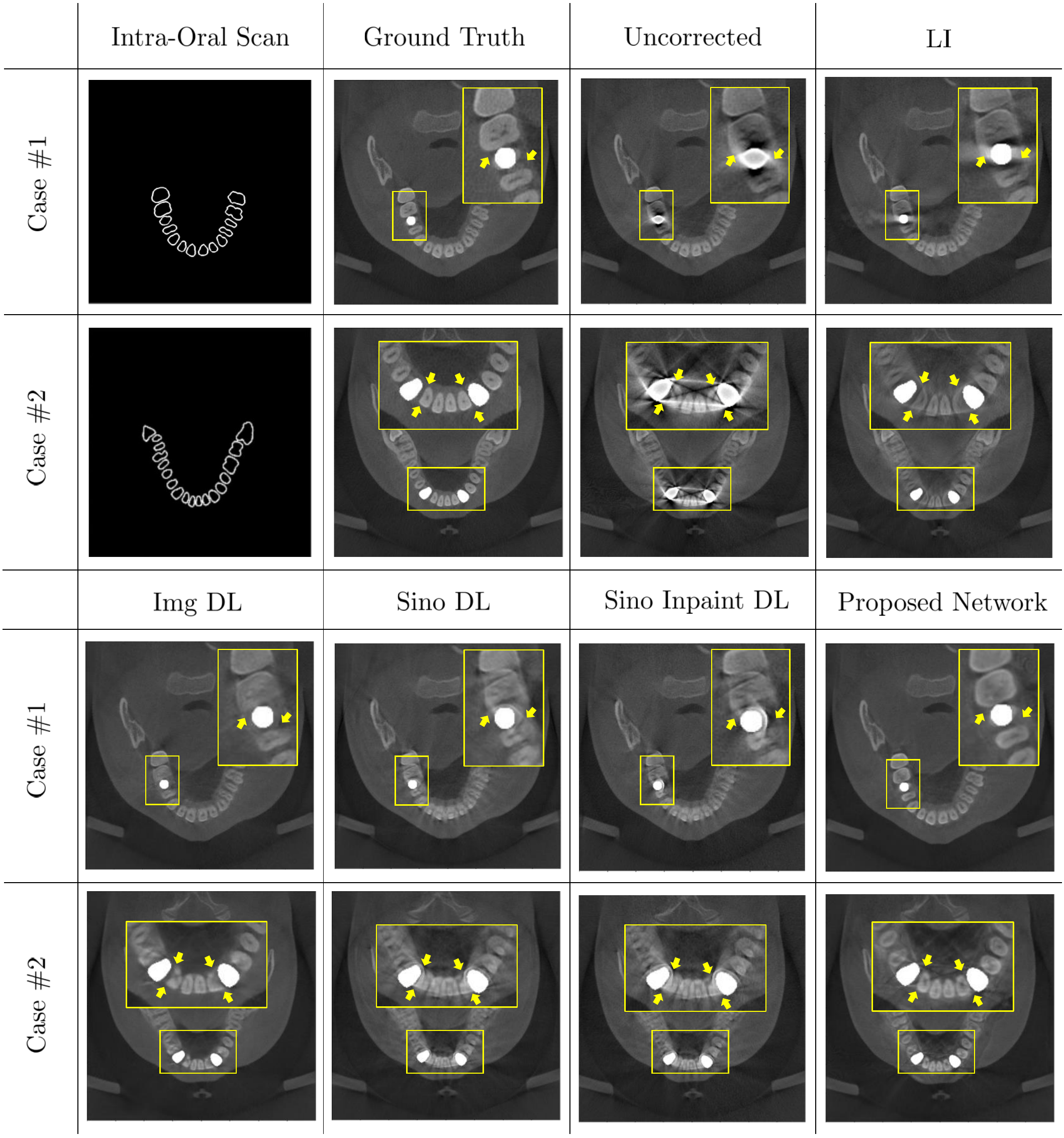}
	\caption{Qualitative comparison of metal artifact reduction over simulated data with various MAR methods; linear interpolation (LI), image domain learning (Img DL), sinogram domain learning (Sino DL), sinogram inpainting learning (Sino Inpaint DL), and the proposed network. Case 1 is the best MAR case and Case 2 is the worst MAR case.}
	\label{exp_result}
\end{figure*}
\begin{figure*}[h]
	\centering
	\includegraphics[width=0.975\textwidth]{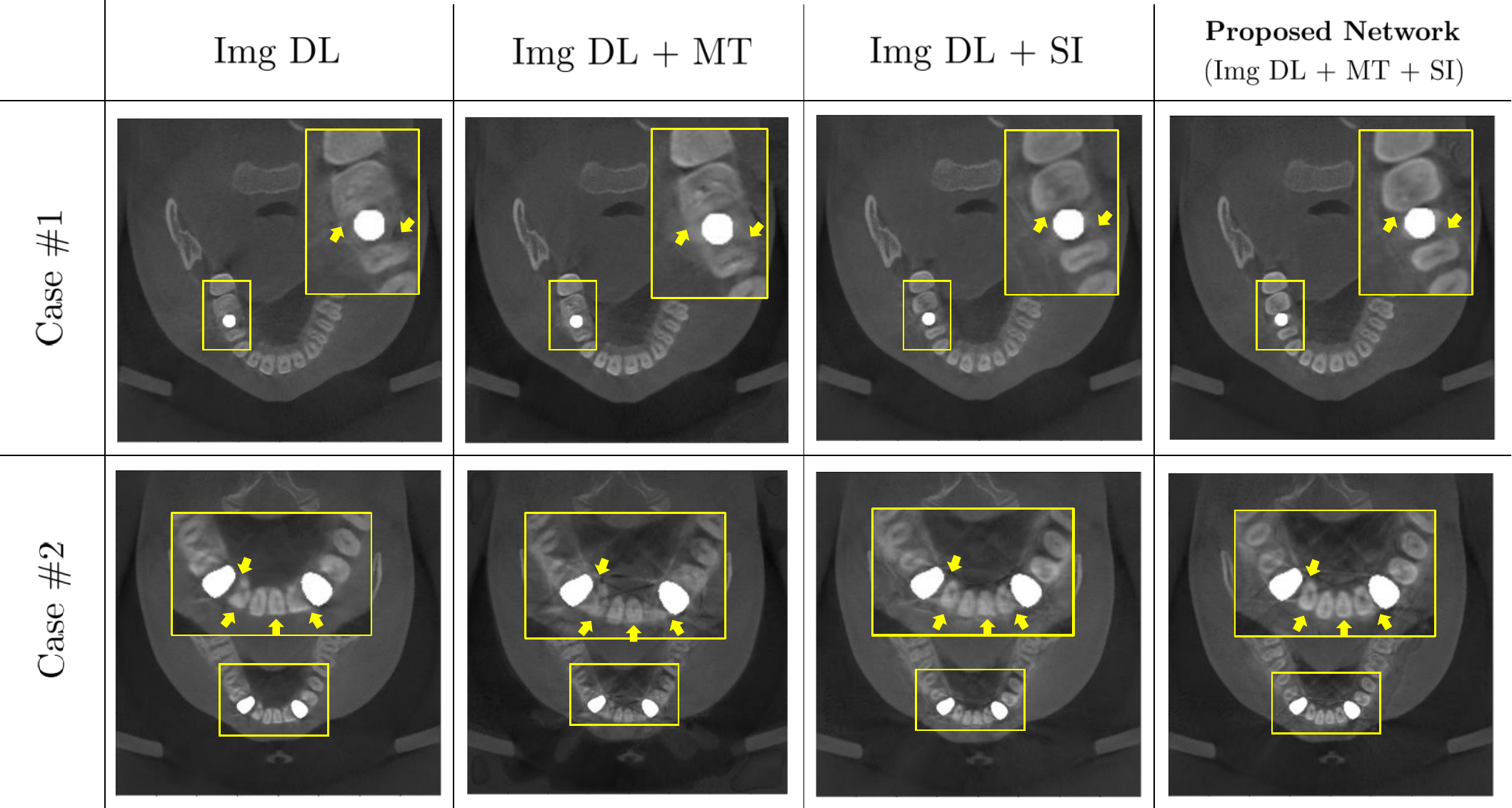}
	\caption{Qualitative ablation study for the proposed method; multitask learning (MT), side input layer (SI).}
	\label{exp_result_sub}
\end{figure*}

Fully automated realistic paired data generation for supervised learning was employed. For given metal-affected data, it is a huge challenge to find the corresponding metal-free data owing to the nonlinear nature of the artifacts. Hence, collecting a clinical paired CBCT dataset of metal-affected and metal-free data from many patients is almost impossible. Instead, a paired dataset can be generated by artificially producing metal artifacts using the Beer-Lambert law \eqref{Beerslaw} with simulated surgery on many normal patient data. However, the generation of realistic data requires time-consuming and labor-intensive manual processes to suitably place dental metallic prostheses within the oral structure. To address this difficulty, a fully-automated tool is proposed that provides realistic shape generation and placement of dental metallic prostheses. The tool is designed for dental crown and tooth implants, which are commonly encountered in clinical dentistry. The overall workflow is shown in Fig. \ref{trainingdatageneration}.

As a first step, fully-automated individual tooth segmentation  was performed on normal patient data by using the technique reported by \cite{jang2021}. Several tooth positions were chosen randomly in which virtual metal implants could be placed. For the crown case, a crown mask was constructed by cutting the roots of chosen teeth based on crown height information for each tooth \cite{nelson2014}, and then by the erosion process. The crown thickness was randomly set from 0.6 to 1.4 mm. For an implant case, instead of erosion, another process was applied to create an implant screw bar. A line was defined for each tooth that passed through two points of the tooth center in the lowest and middle slices, except those containing a tooth root. Then, the root parts were filled with circles whose center was located at the line, and the radius was empirically set. Using the generated dental crown or tooth implant mask, metal-affected sinogram data was artificially synthesized using the Beer-Lambert law \eqref{Beerslaw} and combined with normal patient sinogram.

Simulated intra-oral scan data (i.e., a binary voxel mask of the surface of the teeth) were synthesized by erosion on the combined mask of tooth segmentation and the inserted metal masks.

In the experiment, the same settings were used as those of the real dental CBCT machine: an X-ray source with a tube voltage of 85keV and a tube current of 8mA. A metal attenuation coefficient was randomly assigned from \{Au, Pd, Ni, Cr, Zr, Al\}. For the numerical simulation, the energy distribution of the X-ray source and attenuation coefficient values were those described elsewhere \cite{hubbell1995,mahesh2013}. Poisson and Gaussian noise were added to take account of the CT noise. 
Seventy 3D CBCT data pairs were generated from 20 normal patient data, where the number of inserted metal implants was randomly set from two to five. Fig. \ref{trainingdatageneration2} shows several samples of the simulated data using the data generation tool.

\subsection{Experimental Results}
\begin{table*}[h]
	\centering
	{\footnotesize
		\begin{tabular}{C{1cm}||C{1.5cm}|C{1.5cm}|C{1.5cm}|C{1.5cm}|C{1.5cm}|C{1.5cm}|C{1.5cm}|C{2.2cm}}
			Metric & Uncorrected & LI & Sino DL & Sino Inp DL & Img DL & Img DL+MT & Img DL+SI & Proposed Network \\ \hline \hline
			NMSE & 0.6298  & 0.4158 & 0.5445  & 0.7017 & 0.3577 & 0.3885  & 0.3458  & \textbf{0.3421} \\ \hline
			SSIM & 0.9908  & 0.9948  & 0.9882 & 0.9846  & 0.9954 & 0.9924 & 0.9963  & \textbf{0.9965} \\ \hline
			PSNR & 52.32 & 55.74 & 53.40 & 51.20 & 57.06  & 56.36 & 57.35  & \textbf{57.44}
	\end{tabular}}
	\caption{Quantitative comparison of deep learning-based MAR results for simulated patient data in terms of normalized mean square error (NMSE), structural similarity index (SSIM), and peak signal-to-noise ratio (PSNR).}
	\label{exp_result_table}
\end{table*}

Fig. \ref{exp_result_real_maxillofacial} shows 3D segmented maxillofacial models by uncorrected image + image thresholding, the proposed network + image thresholding, and the proposed method (the proposed network + the $\alpha$-shape-based weighted thresholding). The result was obtained using clinical CBCT data and real intra-oral scan data. The proposed method clearly enhanced the quality of a 3D maxillofacial model so that it precisely depicted the tooth and bone structures. The $\alpha$-shape-based weighted thresholding was found to be powerful in real intra-oral scan data for high quality maxillofacial modeling.

To investigate the advantages of the proposed network, performance comparisons were conducted with various MAR methods. The experiments were based on three test sets: synthesized CBCT data + simulated intra-oral scan data, clinical CBCT data + simulated intra-oral scan data, and clinical CBCT data + real intra-oral scan data. Qualitative and quantitative evaluations were conducted on the synthesized CBCT dataset in which the corresponding ground-truth images are given. For clinical CBCT data, qualitative evaluations were performed.

It should be mentioned that comparison with other methods is unfair, because the proposed method takes advantage of additional information from intra-oral scan data.

\begin{figure*}[t!]
	\centering
	\includegraphics[width=0.975\textwidth]{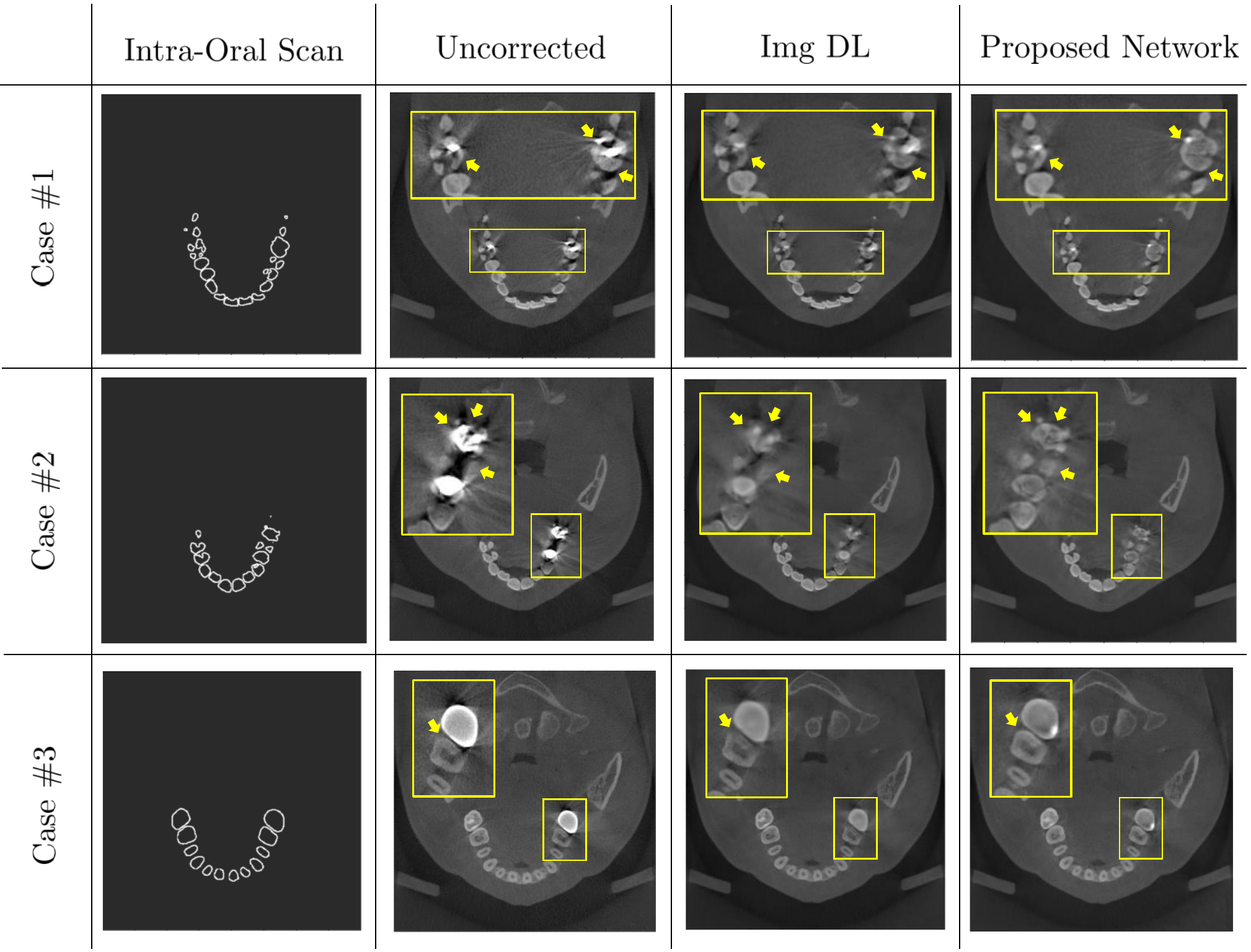}
	\caption{Qualitative comparison of metal artifact reduction with clinical CBCT data and simulated intra-oral scan data; image domain learning (Img DL) and the proposed method.}
	\label{exp_result_realsynt}
\end{figure*}
\begin{figure*}[t!]
	\centering
	\includegraphics[width=0.975\textwidth]{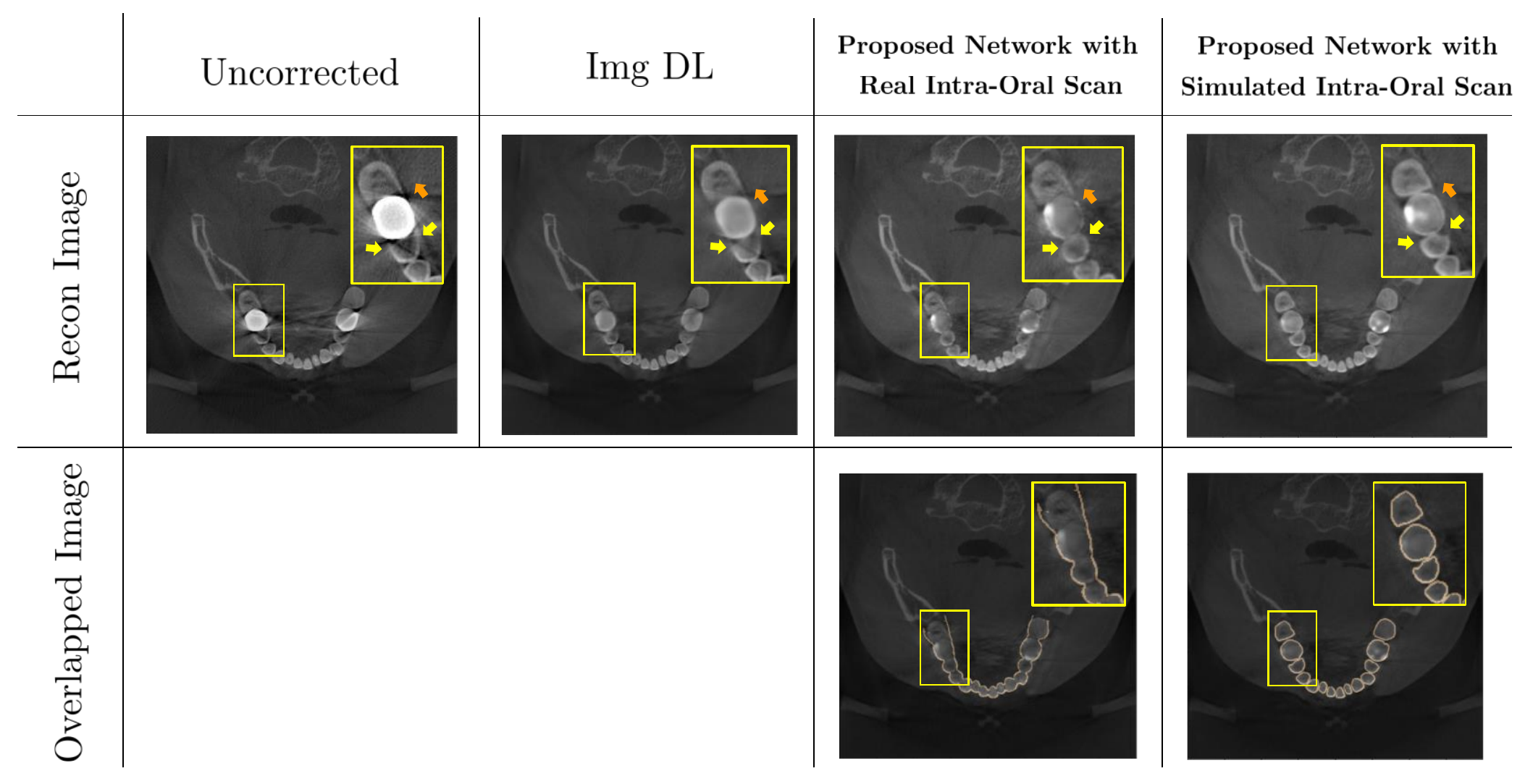}
	\caption{Qualitative comparison of metal artifact reduction with clinical CBCT data; image domain learning (Img DL), the proposed method with real intra-oral scan data, the proposed method with simulated intra-oral scan data. In the second row, we provides an overlapped image of a reconstructed image with the corresponding intra-oral scan data (solid line with apricot color).}
	\label{exp_result_realreal}
\end{figure*}

\subsubsection{Test on synthesized CBCT and simulated intra-oral scan data}
Fig. \ref{exp_result} and Table \ref{exp_result_table} show qualitative and quantitative performance comparisons of the proposed network with linear interpolation, an image domain network, a sinogram domain network, and a sinogram inpainting network. For the linear interpolation, the sinogram reflection technique reported by \cite{bayaraa2020} was applied to deal with metal trace truncation. Image thresholding was used to extract metal traces. For the image domain network, U-net \cite{ronneberger2015} was trained, which directly maps from an uncorrected image to the corresponding ground truth image. For the sinogram domain network, U-net was trained, which directly maps from an uncorrected sinogram to the corresponding ground truth sinogram. For the sinogram inpainting network, U-net was trained such that only the metal traces in the sinogram were corrected by a network output. In the experiments, the proposed network exhibited the best performance, significantly improving the shape quality of teeth and bone associated with maxillofacial imaging. In particular, the proposed network appears to have an outstanding ability to recover the tooth shape, even if it is fairly disrupted or missed because of metal-related artifacts.

As shown in Fig. \ref{exp_result_sub} and Table \ref{exp_result_table}, an ablation study for multi-task learning (MT) and side input layer (SI) in the proposed network was conducted qualitatively and quantitatively. The single use of MT did not provide any advantage in the sense of improving the reconstruction ability in the quantitative and qualitative sense. Either SI or a combination of SI and MT enhances the reconstruction performance both qualitatively and quantitatively. The combination of SI and MT appears to provide an optimal result owing to the synergistic effect.

\subsubsection{Test on clinical CBCT and simulated intra-oral scan data}
Fig. \ref{exp_result_realsynt} shows a qualitative comparison of the test set of real clinical CBCT data and simulated intra-oral scan data, where the intra-oral scan data were obtained by tooth segmentation from the clinical CBCT data. Here, the method of \cite{jang2021} was utilized, which provides considerably accurate tooth segmentation, even in the presence of metal-related artifacts. Several simulated intra-oral scan data are listed in the first column of Fig. \ref{exp_result_realsynt}. In three cases from different patients, the proposed network successfully reduced metal artifacts while recovering the boundary of the teeth effectively, whereas the image domain network tended to suffer from loss, blurring, or disruption of the tooth boundary around metal objects. 

\subsubsection{Test on clinical CBCT and real intra-oral scan data}
Fig. \ref{exp_result_realreal} shows reconstructed results using clinical CBCT and real intra-oral scan data. The proposed method consistently preserves or recovers the boundary of the teeth around metal objects compared with the image domain network. See regions highlighted by yellow arrows in Fig. \ref{exp_result_realreal}.

The performance of the proposed method was compared as well when using simulated and real intra-oral scan data for the same clinical CBCT data. There was some performance degradation in the case of real intra-oral scan data relative to the simulated intra-oral scan case. See the region indicated by the orange arrows in Fig. \ref{exp_result_realreal}.

\section{Conclusion and Discussion} \label{Conclude}
In response to concerns about radiation, dental CBCT has been being developed toward the direction of minimizing radiation exposure while maintaining diagnostic image quality. Scanning using the lowest possible tube current setting and the shortest possible exposure time is recommended. Moreover, most dental CBCTs use a detector with a limited size and an offset arrangement to maximize the axial FOV while reducing the radiation dose or manufacturing cost \cite{chang1995,cho1995}.

In the presence of strong and multiple metallic implants in the FOV of dental CBCT, the measured sinogram is highly contaminated under the influence of complex factors related to low radiation. In particular, the sinograms are frequently missing along the metal trace, resulting in loss of tooth structures around the metallic implants. Moreover, the limited size of a detector and offset arrangement result in significant truncation of the sinogram, as shown in Fig. \ref{intro}. This limited CBCT environment makes it more difficult to apply existing MAR methods.

In this article, an MAR method for high-quality 3D maxillofacial modeling in a low-dose dental CBCT environment was proposed. With the rapid development of digital dentistry technology, it has become possible to obtain 3D surface information about teeth using an intra-oral scanner, which has a huge potential to improve the MAR process. This study propose for the first time a MAR method utilizing an intra-oral scanner for both MAR and maxillofacial modeling in low-dose dental CBCT, which does not require additional radiation exposure for data acquisition. In the experiments, the tremendous potential of intra-oral scan data to have a significant positive effect on the restoration of tooth shape loss caused by low-dose scans was demonstrated. Although there is still much room for improvement, this study is meaningful as a first attempt to pave the way toward MAR utilizing the shape prior from intra-oral scan data.

To train the proposed network, a paired dataset of metal-artifacted CBCT, metal-artifact-free CBCT, and intra-oral scan data is required, but data accessibility is limited in clinical practice. Hence, the data generation tool was utilized to provide realistic metal-artifacted CBCT and intra-oral scan data from metal-free CBCT data, where the intra-oral scan data for training are simulated as a set of boundaries of individual teeth segmented in an artifact-free CBCT image. However, the simulation does not fully reflect the real scanning environment, such as scanning protocol, condition, and performance. The difference between the training and test domains results in the performance degradation of the proposed MAR network, as shown in Fig.  \ref{exp_result_realreal}. The performance of the proposed network on real intra-oral scan data can be improved if more realistic simulated oral scan data or a sufficient number of real oral scan data for training can be obtained.

Moreover, the ability of the learning-based MAR method can be further improved through complex network architectures (e.g., deep layers or large feature depths) and a large-scale paired training dataset. However, there is a trade-off with the total computational cost for learning that can be critical, especially in high-dimensional data applications \cite{hyun2020}. Even for the simple M-net architecture shown in Fig. \ref{proposedmethod}, at least 10 days are required  for training of 300 epochs with a dataset of 60 image voxels under the computational resources used in this study. Even though the use of sophisticated networks or large training datasets can potentially enhance MAR capability, associated hurdles involving high data dimensionality should be addressed for practical dental CBCT applications.

\section*{Acknowledgments}
This work was supported by Samsung Science \& Technology Foundation (No. SRFC-IT1902-09). H.S. Park was partially supported by the National Institute for Mathematical Sciences (NIMS) grant funded by the Korean government (No. NIMS-B22910000). We are deeply grateful to Dr. H. Jung and Dr. S. M. Lee of HDXWILL for their help and collaboration.

\end{document}